\documentclass[%
aps,
prapplied,
reprint,
amsmath,amssymb,
superscriptaddress
]{revtex4-1}

\usepackage{graphicx}% Include figure files
\usepackage{dcolumn}% Align table columns on decimal point
\usepackage{bm}% bold math
\usepackage[mathlines]{lineno}% Enable numbering of text and display math
\usepackage{siunitx}
\usepackage{amsmath}
\usepackage[utf8]{inputenc}
\usepackage[T1]{fontenc}

\usepackage{mathptmx}
\usepackage{etoolbox}

\usepackage{color,soul} % use \hl{highlighting} or \st{overstriking}
\usepackage{comment} 	% Selectively include/exclude portions of text
\usepackage{xcolor} %required for \textcolor command

\makeatletter
\def\@email#1#2{%
 \endgroup
 \patchcmd{\titleblock@produce}
  {\frontmatter@RRAPformat}
  {\frontmatter@RRAPformat{\produce@RRAP{*#1\href{mailto:#2}{#2}}}\frontmatter@RRAPformat}
  {}{}
}%
\makeatother
\begin{document}

%\preprint{AIP/123-QED}

\title{Determining Young's modulus via the eigenmode spectrum of a nanomechanical string resonator}
% Force line breaks with \\

\author{Yannick S. Kla{\ss}}%
\affiliation{%
	Department of Electrical and Computer Engineering, Technical University of Munich, 85748 Garching, Germany
}%
\affiliation{%
	Department of Physics, University of Konstanz, 78457 Konstanz, Germany
}%
	\author{Juliane Doster}
\affiliation{%
	Department of Physics, University of Konstanz, 78457 Konstanz, Germany
}%

	\author{Maximilian B{\"u}ckle}
\affiliation{%
	Department of Physics, University of Konstanz, 78457 Konstanz, Germany
}%

\author{R\'{e}my Braive}
\affiliation{%
	Centre de Nanosciences et de Nanotechnologies, CNRS, Universit\'{e} Paris-Sud, Universit\'{e} Paris-Saclay, 91767 Palaiseau, France
}%
\affiliation{%
	Universit\'{e} de Paris, 75207 Paris Cedex 13, France
}%

\author{Eva M. Weig}
\email{eva.weig@tum.de}

\affiliation{%
	Department of Electrical and Computer Engineering, Technical University of Munich, 85748 Garching, Germany
}%
\affiliation{%
	Department of Physics, University of Konstanz, 78457 Konstanz, Germany
}%
\affiliation{
	Munich Center for Quantum Science and Technology (MCQST), 80799 Munich, Germany
}%
\affiliation{
	TUM Center for Quantum Engineering (ZQE), 85748 Garching, Germany
}%

\date{\today}% It is always \today, today,
             %  but any date may be explicitly specified

\begin{abstract}
We present a method for the in-situ determination of Young's modulus of a nanomechanical string resonator subjected to tensile stress. It relies on measuring a large number of harmonic eigenmodes and allows to access Young's modulus even for the case of a stress-dominated frequency response. We use the proposed framework to obtain Young's modulus of four different wafer materials, comprising the three different material platforms amorphous silicon nitride, crystalline silicon carbide and crystalline indium gallium phosphide. The resulting values are compared with theoretical and literature values where available, revealing the need to measure Young's modulus on the sample material under investigation for precise device characterization.\\

Copyright 2022 Author(s). This article is distributed under a Creative Commons Attribution (CC BY) License.
\end{abstract}

\maketitle

%\section{Introduction}

\begin{figure}[tbh]
	\includegraphics[width=0.95\linewidth]{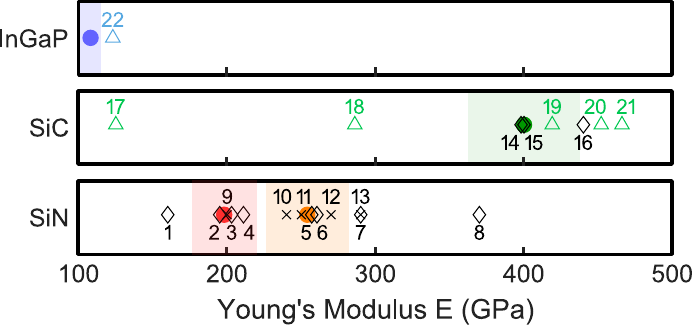}% {ELit.pdf} Here is how to import EPS art
	\caption{
		\label{fig:ELit} 
		Young's modulus for In$_{0.415}$Ga$_{0.585}$P, 3C-SiC, and LPCVD Si$_3$N$_4$. %\ce{Si3N4}.
		Our measured values and uncertainties are shown as filled colored circles and colored shades, respectively, whereas literature values are represented as open symbols. Colored open triangles correspond to values computed form literature values of the elastic constants, matching the crystal direction of the investigated resonators. Measured and other literature values are shown as open black diamonds and crosses, respectively. 
		For the sake of visibility we omit all stated uncertainties. Values are taken from: 1\cite{unterreithmeier2010damping}, 2\cite{guo2001characterization}, 3\cite{zhang2000microbridge}, 4\cite{verbridge2006high}, 5\cite{edwards2004comparison}, 6\cite{chuang2004mechanical}, 7\cite{tabata1989mechanical}, 8\cite{yoshioka2000tensile}, 9\cite{maillet2017nonlinear}, 10\cite{villanueva2014evidence}, 11\cite{Ghadimi2018StrainEngineering,sadeghi2019influence},12\cite{Tsaturyan2017NatNano-UltracoherentNanomechSoftclamping},13\cite{chan2009optical}, 14\cite{iacopi2013evidence}, 15\cite{iacopi2013orientation} , 16\cite{kulikovsky2008hardness}, 17\cite{meyer1926gmelins}, 18\cite{ka2008measurement}, 19\cite{li1987single}, 20\cite{wang1996pressure}, 21\cite{karch1994ab}, 22\cite{Bueckle2018APL-StressControl}. Labels for measured values are found below the corresponding symbol, while all other labels are situated above. 
	}
\end{figure}

Young's modulus of a material determines its stiffness under uniaxial loading. It is thus a crucial  material parameter for many applications involving mechanical or acoustic degrees of freedom, including nano- and micromechanical systems~\cite{Bachtold2022}, cavity optomechanics~\cite{aspelmeyer2014cavity}, surface or bulk acoustic waves including quantum acoustics~\cite{delsing2019,clerk2020hybrid}, nanophononics~\cite{volz2016nanophononics}, or solid-state-based spin mechanics~\cite{perdriat2021spin}, just to name a few. For quantitative prediction or characterization of the performance of those devices, precise knowledge of Young's modulus is required. This is particularly important, as the value of Young's modulus of most materials has been known to strongly depend on growth and even nanofabrication conditions such that relying on literature values may lead to significant deviations~\cite{hahnlein2014mechanical,hahnlein2015mechanical,babaei2009size,kulikovsky2008hardness}. This is apparent from Fig.~\ref{fig:ELit} where we show examples of experimentally and theoretically determined values of Young's modulus along with common literature values for three different material platforms.
For amorphous stoichiometric Si$_3$N$_4$ grown by low pressure chemical vapor deposition (LPCVD), for instance, experimental values between \SI{160}{\giga\pascal}~\cite{unterreithmeier2010damping} and \SI{370}{\giga\pascal}~\cite{yoshioka2000tensile} have been reported. 
The situation is considerably more complex for crystalline materials, for which additional parameters such as the crystal direction or, in the case of polymorphism or polytypism, even the specific crystal structure, affect the elastic properties. 
For these materials, Young's modulus can in principle be calculated via the elastic constants of the crystal~\cite{Bueckle2018APL-StressControl}. However, its determination may be impeded by the lack of literature values of the elastic constants for the crystal structure under investigation, such that the database for theoretical values is scarce. 
This is seen for the ternary semiconductor alloy In$_{1-x}$Ga$_x$P, where even the gallium content $x$ influences Young's modulus~\cite{Bueckle2018APL-StressControl}. 
For 3C-SiC, another crystalline material, theoretical predictions vary between \SI{125}{\giga\pascal}~\cite{meyer1926gmelins} and \SI{466}{\giga\pascal}~\cite{karch1994ab}, even surpassing the spread of experimentally determined values, because literature provides differing values of the elastic constants. 
The apparent spread of the reported values clearly calls for reliable local and in-situ characterization methods applicable to individual devices.

While Young's modulus of macroscopic bulk or thin film samples is conveniently characterized using ultrasonic methods~\cite{guo2001characterization,schneider1996non} or static techniques such as nanoindentation~\cite{iacopi2013orientation}, load deflection~\cite{zhang2000microbridge,yoshioka2000tensile} or bulge testing~\cite{edwards2004comparison,tabata1989mechanical,iacopi2013evidence}, determining it's value on a nanostructure is far from trivial.
%\hl{Load deflection kann/wird auch mit nanostructures gemacht werden.}. 
For freely suspended nanobeams and cantilevers, a dynamical characterization via the eigenfrequency provides reliable results~\cite{hahnlein2014mechanical,hahnlein2015mechanical,chuang2004mechanical,schmid2011damping,verbridge2006high}.
However, this method fails for nanomechanical devices such as membranes or strings subject to a strong intrinsic tensile prestress where the contribution of the bending rigidity and thus Young's modulus to the eigenfrequency becomes negligible. Given the continuously rising interest in this type of materials resulting from the remarkably high mechanical quality factors of several \SI{100000}{} at room temperature~\cite{verbridge2006high,Cole2014APL-InGaP-Membrane,Kermany2014APL-QfactorsOverMillionStressedSiC} arising from dissipation dilution ~\cite{GonzalezSaulson1994BrownianMotionAnelasticWire,Yu2012PRL-ControlMaterialDampingHighQMembraneMicrores,unterreithmeier2010damping}, which can be boosted into the millions by soft clamping and further advanced concepts ~\cite{Tsaturyan2017NatNano-UltracoherentNanomechSoftclamping,Ghadimi2018StrainEngineering}, this calls for an accurate method to determine Young's modulus of stressed nanomechanical resonators. 

Here we present a method for the in-situ determination of Young's modulus of nanomechanical string resonators. It is based on Euler-Bernoulli beam theory and relies on the experimental characterization of a large number of harmonic eigenmodes, which enables us to extract the influence of the bending rigidity on the eigenfrequency despite its minor contribution. We showcase the proposed method to determine the respective Young's modulus of four different wafers, covering all three material platforms outlined in Fig.~\ref{fig:ELit}.  \\

%\section{Theoretical considerations}

According to Euler-Bernoulli beam theory the out-of-plane flexural eigenfrequencies of a doubly clamped string subjected to tensile stress with simply supported boundary conditions are calculated as ~\cite{Timoshenko1990-VibrationProblemsEngineering,Cleland2002foundations}
\begin{equation}
	f_n = \frac{n^2 \pi}{2 L^2}  \sqrt{\frac{E h^2}{12\rho} + \frac{\sigma L^2}{n^2 \pi^2 \rho}} \label{eq:frequency} \,
\end{equation}
where $ n $ is the mode number, $ L $ the length and $ h $ the thickness of the resonator, $ \rho $ the density, $ E $ Young's modulus and $ \sigma $ the tensile stress. For the case of strongly stressed nanostrings, the bending contribution to the eigenfrequency, i.e. the first term unter the square root, will only have a minor contribution compared to the significantly larger stress term. Hence the eigenfrequency-vs.-mode number diagram will approximate the linear behavior of a vibrating string, $f_n \approx (n/2 L) \sqrt{\sigma / \rho}$. So even for a large number of measured harmonic eigenmodes, only minute deviations from linear behavior imply that Young's modulus can only be extracted with a large uncertainty. However, computing $ {f_n^2}/{n^2} $ for two different mode numbers and subtracting them from each other allows to cancel the stress term from the equation, yielding 
\begin{equation}
	\frac{f_n^2}{n^2} - \frac{f_m^2}{m^2} = E \frac{\pi^2 h^2 (n^2 - m^2)}{48 L^4 \varrho}
\end{equation}
with $ m \neq n $. This equation can be solved for Young's modulus
\begin{equation}
	E = \frac{48 L^4 \varrho}{\pi^2 h^2 (n^2 - m^2)} \cdot \left(\frac{f_n^2}{n^2} - \frac{f_m^2}{m^2} \right), \label{eq:YoungsModulus}
\end{equation}
which allows to determine Young's modulus from just the basic dimensions of the string resonator, the density, and the measured eigenfrequency of two different modes. 

The associated uncertainty $\delta E$ obtained by propagation of the uncertainties of all parameters entering  Eq.~\eqref{eq:YoungsModulus} is discussed in the Supplementary Material (SM). We show that the uncertainty of the density, the thickness and the length of the string lead to a constant contribution to $\delta E$ which does not depend on the mode numbers $n$ and $m$. The uncertainty of the eigenfrequencies, however, is minimized for high mode numbers and a large difference between $n$ and $m$.
Therefore it is indispensable to experimentally probe a large number of harmonic eigenmodes to enable a precise result for Young's modulus.\\

\begin{figure}[t!]
	\includegraphics[width=0.9\linewidth]{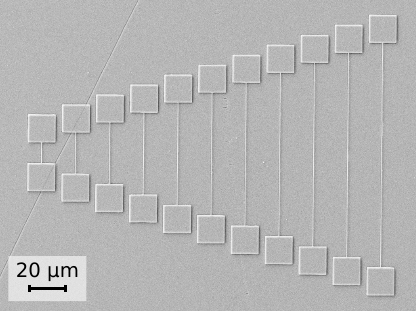}% Here is how to import EPS art
	\caption{
		\label{fig:SEM} 
		Scanning electron micrograph of a series of nano\-string resonators with lengths increasing from \SI{10}{\micro\meter} to \SI{110}{\micro\meter} in steps of \SI{10}{\micro\meter}. 
	}
\end{figure}

%\section{Experimental results}

To validate the proposed method, we are analyzing samples fabricated from four different wafers on the three material platforms outlined in Fig.~\ref{fig:ELit}. Two wafers consist of \SI{100}{\nano\meter} LPCVD-grown amorphous stoichiometric Si$ _3 $N$ _4 $ on a fused silica substrate (denoted as SiN-FS) and on a sacrificial layer of SiO$ _2 $ atop a silicon substrate (SiN-Si), respectively. The third wafer hosts $110$\,nm of epitaxially grown crystalline 3C-SiC on a Si(111) substrate (denoted as SiC). The fourth wafer comprises a \SI{100}{\nano\meter} thick  In$_{0.415}$Ga$_{0.585}$P film epitaxially grown atop a sacrificial layer of Al$_{0.85}$Ga$_{0.15}$As on a GaAs wafer (denoted as InGaP). All four resonator materials exhibit a substantial amount of intrinsic tensile prestress. Details regarding the wafers are listed in the SM.

\begin{figure}[b]
	\includegraphics[width=0.95\linewidth]{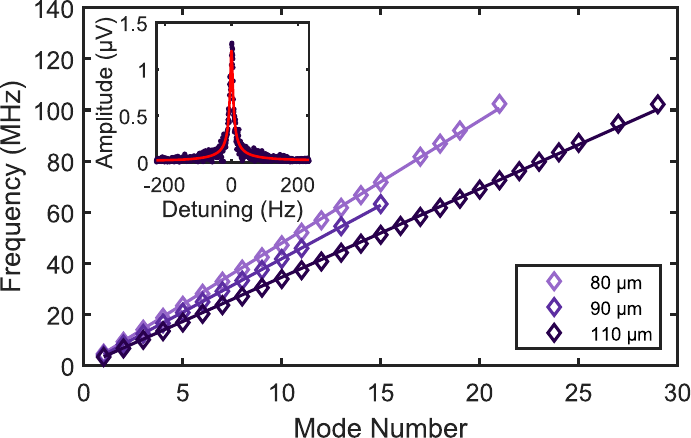}% Here is how to import EPS art
	\caption{
		\label{fig:fvsn} 
		Measured eigenfrequency as a function of the mode number for the three longest SiN-FS strings including fits of the string model (solid lines). Inset depicts the frequency response of the fundamental mode ($ n=1 $, $ L=\SI{110}{\micro\meter} $, $ f_1  = \SI{3.37}{\mega\hertz} $), including a Lorentzian fit (solid lines) to the data (dots).  
	}
\end{figure}

On all four wafers we fabricate series of nanostring resonators with lengths spanning from \SI{10}{\micro\meter} to \SI{110}{\micro\meter} in steps of \SI{10}{\micro\meter} as shown in Fig.~\ref{fig:SEM}. However, as the tensile stress has been shown to depend on the length of the nanostring in a previous work~\cite{buckle2021universal} and might have an impact of Young's modulus~\cite{hong1990measuring}, we focus solely on the three longest strings of each sample for which the tensile stress has converged to a constant value~\cite{buckle2021universal} (see SM for a comparison of Young's modulus of all string lengths). 
	
For each resonator we determine the frequency response for a series of higher harmonics by using piezo actuation and  optical interferometric detection. The drive strength is adjusted to make sure to remain in the linear response regime of each mode. The interferometer operates at a wavelength of $1550$\,nm and is attenuated to operate at the minimal laser power required to obtain a good signal-to-noise ratio to avoid unwanted eigenfrequency shifts caused by absorption-induced heating of the device. This is particularly important as the position of the laser spot has to be adapted to appropriately capture all even and odd harmonic eigenmodes. We extract the resonance frequencies by fitting each mode with a Lorentzian function as visualized in the inset of Fig.~\ref{fig:fvsn}. Figure~\ref{fig:fvsn} depicts the frequency of up to 29 eigenmodes of three SiN-FS string resonators. Solid lines represent fits of the string model ($f_n \approx (n/2 L) \sqrt{\sigma / \rho}$) with $ \sigma $ being the only free parameter (see SM). The slight deviation observed for high mode numbers is a consequence of the bending contribution neglected in this approximation. Note the fit of the full model (Eq.~\eqref{eq:frequency}) yields a somewhat better agreement, however, Young's modulus can not be reliably extracted as a second free parameter in the stress-dominated regime.

However, taking advantage of Eq.~\eqref{eq:YoungsModulus} we can now determine Young's modulus along with its uncertainty for each combination of $n$ and $m$. 
All input parameters as well as their uncertainties are listed in the SM.
To get as much statistics as possible, we introduce the difference of two mode numbers $ \Delta = \left| m-n \right|$ as a parameter. For instance, $ \Delta = 5 $ corresponds to the combinations $ (n=1, m=6) $, $ (2,7) $, $ (3,8) $, $ \ldots $. For each $ \Delta $ we calculate the mean value of $ \overline{E} $ and $ \overline{\delta E}$, respectively. 

\begin{figure}[b]
	\includegraphics[width=0.96\linewidth]{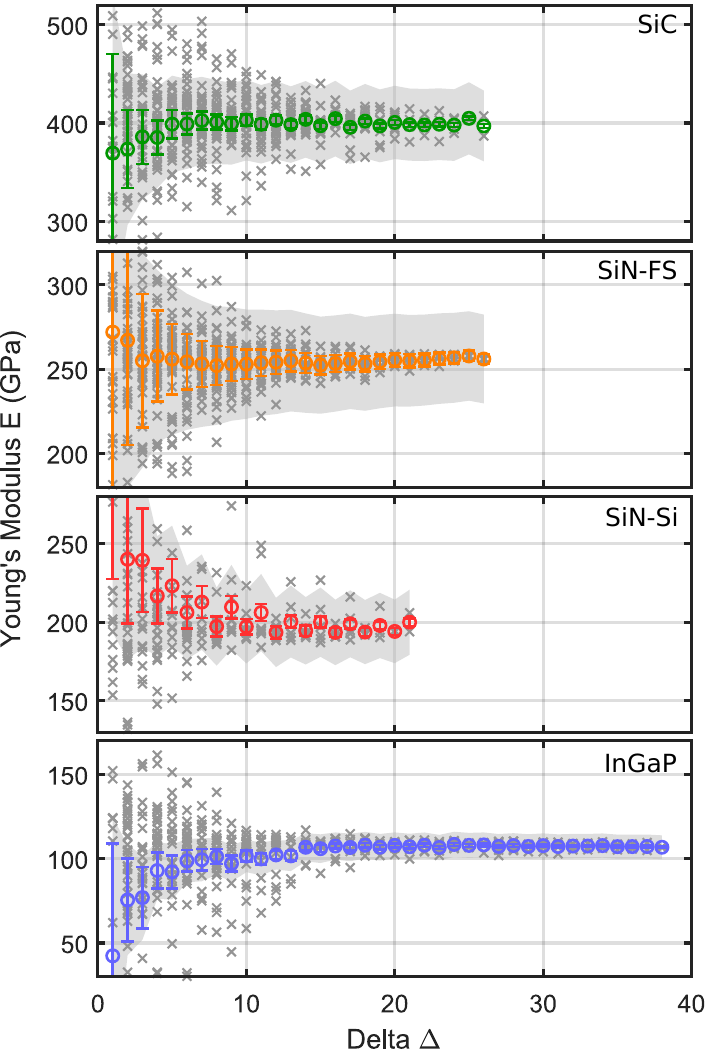}% Here is how to import EPS art
	\caption{
		\label{fig:EAllMat} 
		Determined Young's modulus as a function of $ \Delta $ for the four different materials SiC (green), SiN-FS (orange), SiN-Si (red), and InGaP (blue). Gray crosses correspond to individual combinations of $\left| m-n \right|$. Their mean values $\overline{E}(\Delta)$ are shown as colored circles. Note that while all combinations of $ n $, $ m $ are included in the calculation of $ \overline{E} $ for a given $ \Delta $, not all of them are shown as gray crosses as some heavy outliers appearing mostly for low values of $ \Delta $ have been truncated for the sake of visibility. The complete uncertainty is represented by the gray shade, whereas its $\Delta$-dependent contribution arising from the uncertainty in the eigenfrequency determination is represented by the colored error bars.
	}
\end{figure}

The obtained values of Young's modulus are depicted as a function of $ \Delta $ for all four materials in Fig.~\ref{fig:EAllMat}. Note that only $ \Delta $ values comprising two or more combinations of mode numbers are shown. The individual combinations $ E(\Delta)$ contributing to $ \overline{E} $ for a specific $ \Delta $ are visualized as gray crosses, whereas the mean values of Young's modulus $\overline{E}$ for each value of $ \Delta $ are included as colored circles.

Clearly, Young's modulus of each material converges to a specific value for increasing $ \Delta $. These values are extracted by averaging over the obtained values of $ \overline{E} $ and summarized in Tab.~\ref{tab:results}. 
Note that only the upper half of the available $ \Delta $ points have been included in the average in order to avoid some systematic distortions appearing for low $ \Delta $.

The uncertainty associated with the mean Young's modulus $\overline{\delta E}$ is indicated by gray shades. 
As discussed in more detail in the SM, the $\Delta$-dependence of the uncertainty arises solely from the uncertainty in the eigenfrequency determination. Therefore, this contribution to the total uncertainty is highlighted separately as colored error bars.

For small $ \Delta $, a large uncertainty in the eigenfrequency determination is observed which dominates the complete uncertainty $\overline{\delta E}$. It coincides with a considerable scatter of the individual combinations, which is also attributed to the impact of the eigenfrequency determination.
As expected, for increasing $ \Delta $, the uncertainty in the eigenfrequency determination decreases, such that the complete uncertainty $\overline{\delta E}$ becomes dominated by the constant contribution originating from the uncertainties in the density, thickness and length of the string. The total uncertainty is obtained by averaging $\overline{\delta E}$ over the upper half of the available $ \Delta $ points. It is also included in Tab.~\ref{tab:results}.\\

\begin{table}
	\caption{
		\label{tab:results}
		Young's modulus including the total uncertainty determined for the four different materials.
	}
	\begin{ruledtabular}
		\begin{tabular}{lcccc}
			& SiN-FS & SiN-Si & SiC & InGaP \\ \hline
			$ E $ (GPa) & 254(28) & 198(22) & 400(38) & 108(7) 
			%Fehler für delta f: 5 ,13 ,13, 2
		\end{tabular}
	\end{ruledtabular}
	
\end{table}

%\section{Discussion}

The resulting values for Young's modulus are also included in Fig.~\ref{fig:ELit} as colored dots, using the same color code as in  Fig.~\ref{fig:EAllMat}. Clearly, the determined values coincide with the parameter corridor suggested by our analysis of the existing literature:
For InGaP, where no independent literature values are available we compute Young's modulus~\cite{Bueckle2018APL-StressControl} from the elastic constants of InGaP with the appropriate Ga content ($x=0.585$) and crystal orientation ([110]), yielding $ E_{\textnormal{InGaP}}^{\textnormal{th}} =\SI{123}{\giga\pascal}$~\cite{Ioffe1999ShurEtAl-HandbookSeriesSemiconductorParametersVOL2,Bueckle2018APL-StressControl} (which is included as the theory value for InGaP in Fig.~\ref{fig:EAllMat}). This is rather close to our experimentally determined value of $ E_{\textnormal{InGaP}} =\SI{108\pm7}{\giga\pascal}$. 
For SiC we can calculate Young's modulus as well, however, the elastic constants required for the calculations vary dramatically in literature. As also included as theory values in Fig.~\ref{fig:ELit}, we can produce values of $ E_{\textnormal{SiC}}^{\textnormal{th}} = $ \SI{125}{\giga\pascal}~\cite{meyer1926gmelins}, \SI{286}{\giga\pascal}~\cite{ka2008measurement}, \SI{419}{\giga\pascal}~\cite{li1987single}, \SI{452}{\giga\pascal}~\cite{wang1996pressure}, or \SI{466}{\giga\pascal}~\cite{karch1994ab}, just by choosing different references for the elastic constants. For our material we measure a Young's modulus of $ E_{\textnormal{SiC}} =\SI{400\pm36}{\giga\pascal}$ which is in perfect agreement with the experimentally determined literature values of  \SI{398}{\giga\pascal}~\cite{iacopi2013evidence} and \SI{400}{\giga\pascal}~\cite{iacopi2013orientation} by Iacopi et al.. It also is in good agreement with the elastic constants published by Li and Bradt~\cite{li1987single}, yielding \SI{419}{\giga\pascal} for the orientation of our string resonators.  
Interestingly, SiN-FS and SiN-Si exhibit significantly different Young's moduli of $ E_{\textnormal{SiN-FS}} =\SI{254\pm26}{\giga\pascal}$ and $ E_{\textnormal{SiN-Si}} =\SI{198\pm21}{\giga\pascal}$, respectively. In Fig.~\ref{fig:ELit} we can see two small clusters of measured Young's moduli around our determined values, suggesting that the exact Young's modulus depends on growth conditions and the subjacent substrate material even for the case of an amorphous resonator material. 

In conclusion, we have presented a thorough analysis of Young's modulus of strongly stressed nanostring resonators fabricated from four different wafer materials. The demonstrated method to extract Young's modulus yields an accurate prediction with a well-defined uncertainty. It is suitable for all types of nano- or micromechanical resonators subjected to intrinsic tensile stress. As we also show that literature values provide hardly the required level of accuracy for quantitative analysis, even when considering the appropriate material specifications, the in-situ determination of Young's modulus is an indispensable tool for the precise and complete sample characterization, which can significantly improve the design of nanomechanical devices to fulfill quantitative specifications, or the comparison of experimental data to quantitative models when not using free fitting parameters. Furthermore, the presented strategy can also be applied to two-dimensional tensioned membrane resonators. However, in case of an anisotropic Young's modulus only an average value will be accessible, such that the present case of a one-dimensional string resonator is better suited to characterize Young's modulus of a crystalline resonator.
\\

See supplementary material for a list of used material parameters, a discussion of the uncertainties, and the stress dependence of Young's modulus. \\

Financial support from the European
Unions Horizon 2020 programme for Research and Innovation
under Grant Agreement No. 732894 (FET Proactive HOT) and the German Federal Ministry of Education
and Research through Contract No. 13N14777 funded within
the European QuantERA cofund project QuaSeRT is gratefully
acknowledged. 
We further acknowledge financial support from the Deutsche Forschungsgemeinschaft (DFG, German Research Foundation) through Project-ID 425217212 - SFB 1432 and via project WE 4721/1-1, as well as project QT-6 SPOC of the Baden-W\"urttemberg Foundation.\\

The data and scripts that support the findings of this study are openly available on zenodo (http://doi.org/10.5281/zenodo.6951670).

%\nocite{*}
\bibliography{main}% Produces the bibliography via BibTeX.

\end{document}

\title{Supplemental Material: Determining Young's modulus via the eigenmode spectrum of a nanomechanical string resonator}% Force line breaks with \\
	%\thanks{A footnote to the article title}%
	
	\author{Yannick S. Kla{\ss}}%
	\affiliation{%
		Department of Electrical and Computer Engineering, Technical University of Munich, 85748 Garching, Germany
	}%
	\affiliation{%
		Department of Physics, University of Konstanz, 78457 Konstanz, Germany
	}%
	\author{Juliane Doster}
	\affiliation{%
		Department of Physics, University of Konstanz, 78457 Konstanz, Germany
	}%
	
	\author{Maximilian B{\"u}ckle}
	\affiliation{%
		Department of Physics, University of Konstanz, 78457 Konstanz, Germany
	}%

	\author{R\'{e}my Braive}
	\affiliation{%
		Centre de Nanosciences et de Nanotechnologies, CNRS, Universit\'{e} Paris-Sud, Universit\'{e} Paris-Saclay, 91767 Palaiseau, France
	}%
	\affiliation{%
		Universit\'{e} de Paris, 75207 Paris Cedex 13, France
	}%
	
	\author{Eva M. Weig}
	\email{eva.weig@tum.de}
	
	\affiliation{%
		Department of Electrical and Computer Engineering, Technical University of Munich, 85748 Garching, Germany
	}%
	\affiliation{%
		Department of Physics, University of Konstanz, 78457 Konstanz, Germany
	}%
	\affiliation{
		Munich Center for Quantum Science and Technology (MCQST), 80799 Munich, Germany
	}
	\affiliation{
		TUM Center for Quantum Engineering (ZQE), 85748 Garching, Germany
	}%

	\maketitle

\appendix

	This Supplemental Material provides further information about material parameters in Appendix~\ref{A}, the propagation of uncertainty in Appendix~\ref{B}, and a short discussion of stress dependence of Young's modulus in Appendix~\ref{C}.

\newpage

\section{Material parameters}	
 \label{A}

Table~\ref{tab:wafers} summarizes the growth parameters of the four wafers employed in this work, stating the thickness of the device layer (according to the manufacturer), sacrificial layer (if the system has one), substrate, and the corresponding supplier. The two SiN wafers were grown by Low Pressure Chemical Vapor Deposition (LPCVD), the SiC in a two-stage Chemical Vapor Deposition process, and the In$_{1-x}$Ga$_x$P using Metal-Organic Chemical Vapor Deposition (MOCVD).

\begin{table*}[h!]
	\caption{
		\label{tab:wafers}
		Basic parameters of the wafers on which the string resonators were fabricated. The film thickness is provided by the manufacturer.}
	\begin{ruledtabular}
		\begin{tabular}{lllll}
			& resonator/device layer & sacrificial layer & substrate & source \\
			\hline
			SiN-FS 			& 100(2)\,nm SiN	& --- 	& SiO$_2$ & HSG-IMIT  \\
			SiN-Si 			& 100(2)\,nm SiN 	& 400\,nm SiO$_2$ 	& Si(100) & HSG-IMIT  \\
			SiC 			& 110(2)\,nm 3C-SiC 	& --- 	& Si(111) & NOVASiC  \\
			In$_{1-x}$Ga$_x$P 	& \multicolumn{1}{r}{100(1)\,nm In$_{0.415}$Ga$_{0.585}$P} 	& \multicolumn{1}{l}{1000\,nm Al$_{0.85}$Ga$_{0.15}$As} 	& GaAs & CNRS  \\
		\end{tabular}
	\end{ruledtabular}
\end{table*}

Table~\ref{tab:strings} lists the dimensions and material parameters of the strings required for the Euler-Bernoulli analysis discussed in the main text. The length of the strings has been determined by scanning electron microscopic imaging to incorporate fabricational deviations from the nominal values. The relatively large error is a result of the low magnification required to measure the length of the longest string. Note that with the exception of the SiC sample, a few of the longest strings were broken, such that the three longest available strings on each sample were investigated. The thickness of the strings has not been measured separately; we assume the thickness of the device layer (see Tab.~1). The error is estimated from manufacturer specifications. The mass density of the materials is taken from literature. We would like to emphasize that these values, like those for Young's modulus, also exhibit a substantial spread for all three materials under investigation. We chose one of the intermediate values and assumed a large error to account for that uncertainty. Finally, the error in determining the eigenfrequency $ \delta f $ is \emph{not} the mere fitting error (which is negligible, see inset of Fig.~3 in main text). It also takes into account frequency shifts due to temperature fluctuations over the course of the measurement. These are not only attributed to fluctuations in the ambient temperature, but can also be caused by an adjustment of the position of the laser spot which was required to capture all even and odd harmonic eigenmodes.

\begin{table*}[h!]
	\caption{
		\label{tab:strings}
		Parameters used for the calculations in the main text. It includes the measured length of the strings, the density, and the uncertainty of the frequency. 
	}
	\begin{ruledtabular}
		\begin{tabular}{lllll}
			& SiN-FS & SiN-Si & SiC & InGaP \\
			\hline
$ h~(\SI{}{\nano\meter})$ & 100(2) & 100(2) & 110(2) & 100(1) \\
$ L $ string 1$~(\SI{}{\micro\meter})$ & 107.7(5)	& 100.4(5) & 109.9(5) & 110.5(5)  \\
$ L $ string 2$~(\SI{}{\micro\meter})$ & 88.0(5)	& 90.3(5)	& 99.9(5) & 90.5(5)	\\
$ L $ string 3$~(\SI{}{\micro\meter})$ & 78.2(5)	& 70.2(5)	& 89.8(5) & 80.5(5) \\
$ \rho~(\SI{}{\gram\per\centi\meter^3}) $ & 3.1(1)\cite{maluf2002introduction,sze2021physics,zhang1992silicon} & 3.1(1)\cite{maluf2002introduction,sze2021physics,zhang1992silicon} & 3.2(1)\cite{henisch2013silicon,harris1995properties} & 4.4(1)\cite{Ioffe1999ShurEtAl-HandbookSeriesSemiconductorParametersVOL2} \\
$ \delta f~(\permil)$ & 0.25 & 0.25 & 0.25 & 0.25 \\
		\end{tabular}
	\end{ruledtabular}

\end{table*}

\clearpage
\newpage

\section{Propagation of uncertainty}
\label{B}

Assuming uncertainties of the eigenfrequency $ \delta f $, length $ \delta L $, thickness $ \delta h $, and density $ \delta \rho $ for the calculation of Young's modulus (Eq.~(3) of the main text), the propagation of uncertainty yields
\begin{align}
	\delta E =& \frac{48 L^4 \varrho}{\pi^2 h^2} \left(  \left| \frac{2 f_n}{n^2 (n^2-m^2)} \right| \cdot \delta f_n + \left|\frac{2 f_m}{m^2 (n^2 - m^2)}\right| \cdot \delta f_m  \right. \notag \\
	& \qquad + \left. \left|\frac{f_n^2}{n^2 (n^2 - m^2)} - \frac{f_m^2}{m^2 (n^2 - m^2)} \right| \left[ \frac{4}{L} \delta L + \frac{2}{h} \delta h + \frac{1}{\rho} \delta \rho \right]  \right). \label{eq:YmdE} 	
\end{align}
This enables us to determine the complete error $ \delta E $ for each combination of $n$ and $m$ by inserting the measured frequencies $ f_{\textnormal{n,m}} $. The gray shade in Fig.~4 of the main text as well as the uncertainties in Tab.~I of the main text correspond to the total uncertainty, i.e. the mean of the uncertainties according to Eq.~(\ref{eq:YmdE}) for each combination of $n$ and $m$ contributing to a particular $ \Delta = \left| m-n \right|$. 
	
\begin{figure}[b]
	\includegraphics[width=0.95\linewidth]{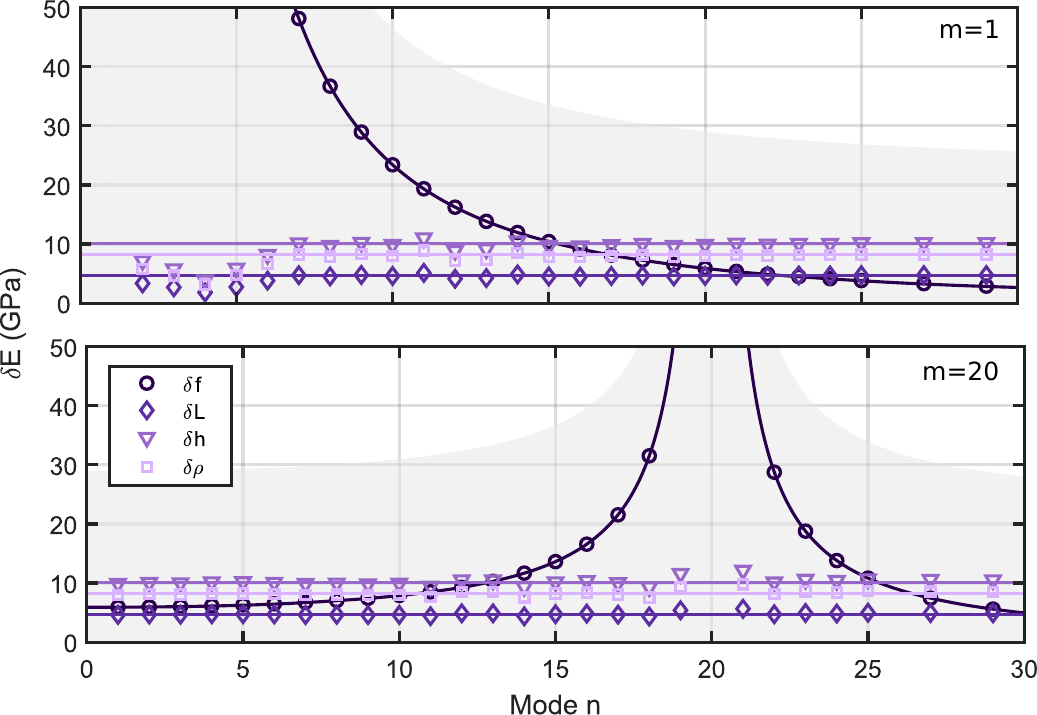}% Here is how to import EPS art
	\caption{
		\label{fig:dEvsN} 
		Contributions of the uncertainties of the input parameters to the uncertainty of Young's modulus for the $ \SI{110}{\micro\meter} $ long SiN-FS string for two fixed mode numbers $ m = 1 $ (top) and $ m=20 $ (bottom). The colored symbols are calculated with Eq.~\eqref{eq:YmdE} and the measured frequencies $ f_{\textnormal{n,m}} $. The colored solid lines are computed with Eq.~\eqref{eq:YmdE2} where we used the values of $ f_{\textnormal{n,m}} $ predicted by Euler Bernoulli beam theory. The gray shaded area indicates the complete error, i.e. the sum of all four contributions.
	}
\end{figure}

For a more detailed discussion, we will now analyze the four contributions to Eq.~(\ref{eq:YmdE}), associated to the uncertainties of the eigenfrequency $ \delta f$, the length $ \delta L$, the thickness $ \delta h$ and the density $ \delta \rho$, individually. They are plotted in Fig.~\ref{fig:dEvsN} for the case of the $110\,\mu$m long SiN-FS string as a function of the mode number $ n $ for the case of $ m = 1 $ and $ 20 $, respectively. 

In order to interpolate between the integer values of $ n $, we insert the eigenfrequencies of the Euler-Bernoulli beam (see Eq.~(1) of the main text) and obtain the simplified expression
\begin{align}
\delta E=& \frac{48 L^4 \varrho}{\pi^2 h^2} \left(  \left| \frac{2 f_n}{n^2 (n^2-m^2)} \right| \cdot \delta f_n + \left|\frac{2 f_m}{m^2 (n^2 - m^2)}\right| \cdot \delta f_m  \right) \notag \\
& \qquad + \frac{4E}{L} \delta L + \frac{2 E}{h} \delta h + \frac{E}{\rho} \delta \rho,  \label{eq:YmdE2} 
\end{align}
the contributions of which are also plotted in Fig.~\ref{fig:dEvsN} as solid lines. Clearly, only the contribution of the eigenfrequency uncertainty depends on the mode numbers $n$ and $m$ (resp. their difference $\Delta$), whereas the contributions of the other three uncertainties are constant. For small $ n $ and for $ n \approx m $, the eigenfrequency error provides the dominant contribution to $ \delta E $. For all other values of $ n $, $ \delta E $ is dominated by the constant uncertainties of the resonator thickness, the density, and the length, whereas the contribution of the eigenfrequency uncertainty becomes negligible. 
Therefore, a precise determination of Young's modulus calls for the experimental determination of a large number of harmonic eigenmodes, in order to perform the analysis with large $ \Delta = \left| m-n \right|$.

To visualize this important observation, the contribution of the uncertainty of the eigenfrequency determination, i.e. the first line of Eq.~(\ref{eq:YmdE}), is included in Fig.~4 of the main text as colored error bars.

All in all, the complete error $\delta E$ is minimized for high mode numbers and a large difference between $ n $ and $ m $ which corresponds to a large value of $ \Delta = \left| m-n \right|$. The complete error, i.e. the sum of all four contributions, is included in Fig.~\ref{fig:dEvsN} as a gray shaded area. Comparison of the two cases, $m=1$ and $m=20$, reveals that the uncertainty obtained for $ m = 1 $ provides the largest contribution to the uncertainty $ \delta E $.

\clearpage
\newpage

\section{Young's modulus as a function of tensile stress} \label{C}

\begin{figure}[b]
	\includegraphics[width=0.95\linewidth]{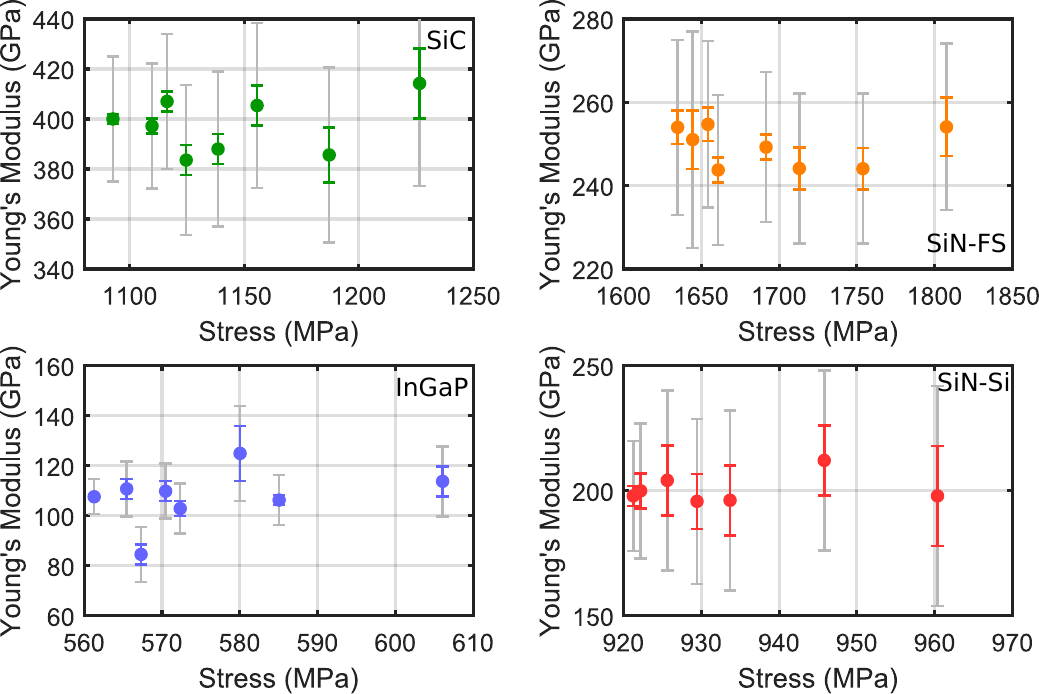}% Here is how to import EPS art
	\caption{
		\label{fig:YmVsStress} 
	Young's Modulus as a function of the tensile stress for four different materials. The dots are determined with the help of Eq.~(3) of the main text. The gray and colored error bars correspond to the full uncertainty (Eq.~\ref{eq:YmdE}) and the uncertainty arising form the frequency (first term in Eq.~\ref{eq:YmdE}). The stress of the individual strings are listed in Tab.~\ref{tab:stress}. 
	}
\end{figure}

Table~\ref{tab:stress} displays the stress of the string resonators used in Fig.~\ref{fig:YmVsStress} and the main text. We extract the stress by fitting Eq.~(1) to the eigenmode spectrum of fixed length $ L $ with $ \sigma $ being the only free parameter. By determining Young's modulus via Eq.~\ref{eq:YmdE} for every single resonator listed in Tab.~\ref{tab:stress}, we are able to plot Young's modulus as a function of tensile stress, as shown in Fig.~\ref{fig:YmVsStress}. Young's modulus of all four materials appears to be approximately constant for all considered tensile stresses. Note that for the shorter strings we could measure significantly less modes, leading to a higher uncertainty, as explained in Sec.~\ref{B}. Equation~(3) of the main text allows us to determine Young's modulus for multiple strings with a different tensile stress as shown in Fig.~\ref{fig:YmVsStress}.

\begin{table*}[h!]
	\caption{
		\label{tab:stress}
		Tensile stress of the string resonators used in Fig.~\ref{fig:YmVsStress} and the main text (three longest string, bold values) extracted via a fit of Eq.~(1) to the eigenmode spectrum with the stress being the only free fit parameter. For the other parameters we used the values given in Tabs.~\ref{tab:strings} and I.  
	}
	\begin{ruledtabular}
		\begin{tabular}{ccccc}
			& SiN-FS & SiN-Si & SiC & InGaP \\
			\hline
			\SI{110}{\micro\meter}: stress $ \sigma $ $~(\SI{}{\mega\pascal})$ & \textbf{1635}	& -- & \textbf{1093} & \textbf{561}  \\
			\SI{100}{\micro\meter}: stress $ \sigma $ $~(\SI{}{\mega\pascal})$ & --	& \textbf{921}	& \textbf{1100} & --	\\
			\SI{90}{\micro\meter}: stress $ \sigma $ $~(\SI{}{\mega\pascal})$ & \textbf{1644}	& \textbf{922}	& \textbf{1110}  & \textbf{565} \\
			\SI{80}{\micro\meter}: stress $ \sigma $ $~(\SI{}{\mega\pascal})$ & \textbf{1654}	& -- & -- & \textbf{567}  \\
			\SI{70}{\micro\meter}: stress $ \sigma $ $~(\SI{}{\mega\pascal})$ & 1661	& \textbf{926}	& 1116 & 570	\\
			\SI{60}{\micro\meter}: stress $ \sigma $ $~(\SI{}{\mega\pascal})$ & 1692	& 929	& 1125  & 572 \\
			\SI{50}{\micro\meter}: stress $ \sigma $ $~(\SI{}{\mega\pascal})$ & 1754	& 934	& 1139  & 580 \\
			\SI{40}{\micro\meter}: stress $ \sigma $ $~(\SI{}{\mega\pascal})$ & 1754	& 946	& 1156  & 585 \\
			\SI{30}{\micro\meter}: stress $ \sigma $ $~(\SI{}{\mega\pascal})$ & 1808	& 960	& 1227  & 606 \\
		\end{tabular}
	\end{ruledtabular}
	
\end{table*}

\clearpage

\bibliography{si}